\documentclass[aps,pre,twocolumn,groupedaddress]{revtex4-1}

\usepackage[utf8]{inputenc}
\usepackage{graphicx,psfrag,color}
\usepackage{caption}
\usepackage{subcaption}
\usepackage{epstopdf}
\usepackage{amsmath}
\usepackage{amsfonts}
\usepackage{mathrsfs}
\usepackage{amssymb}
\usepackage{bm}
\usepackage{url}
\usepackage{color}
\usepackage{natbib}
\usepackage{mdwlist}

\newcommand{\bnab}{\mbox{\boldmath$\nabla$}}

\def\half {{\textstyle{1 \over 2}}}

\newcommand{\bhx}{\mathbf{\hat{x}}}
\newcommand{\bhy}{\mathbf{\hat{y}}}
\newcommand{\bhz}{\mathbf{\hat{z}}}
\newcommand{\bphi}{\mbox{\boldmath$\hat{\theta}$}}
\newcommand{\bOmega}{\mathbf{\Omega}}
\newcommand{\bu}{\mathbf{u}}
\newcommand{\bU}{\mathbf{U}}
\newcommand{\newU}{\phi}

\newcommand{\beq}{\begin{equation}}
\newcommand{\eeq}{\end{equation}}

\graphicspath{{figures/}}

\begin{document}



\title{Connections between centrifugal, stratorotational and radiative instabilities in viscous Taylor--Couette flow}


\author{Colin Leclercq$^1$, Florian Nguyen$^{1,2}$, Rich R. Kerswell$^1$}
\affiliation{$^1$Department of Mathematics, University of Bristol, University Walk, Bristol BS8 1TW, United Kingdom\\
             $^2$\'{E}cole Normale Sup\'{e}rieure, Paris, France}


\date{\today}

\begin{abstract}
The `Rayleigh line' $\mu=\eta^2$, where $\mu=\Omega_o/\Omega_i$ and $\eta=r_i/r_o$ are respectively the rotation and radius ratios between inner (subscript `$i$') and outer (subscript `$o$') cylinders, is regarded as marking the limit of centrifugal instability (CI) in unstratified inviscid Taylor--Couette flow, for both axisymmetric and non-axisymmetric modes.
Non-axisymmetric stratorotational instability (SRI) is known to set in for anticyclonic rotation ratios beyond that line,
 i.e. $\eta^2<\mu<1$ for axially stably-stratified Taylor--Couette flow, but the competition between CI and SRI in the
 range $\mu<\eta^2$ has not yet been addressed. 
In this paper, we establish continuous connections between the two instabilities at finite Reynolds number $Re$, as previously suggested by M. Le Bars \& P. Le Gal, \textit{Phys. Rev. Lett.} \textbf{99}, 064502 (2007), making them indistinguishable at onset.
Both instabilities are also continuously connected to the radiative instability at finite $Re$. 
These results demonstrate the complex impact viscosity has on the linear stability properties of this flow. Several other qualitative differences with inviscid theory were found, among which the instability of a non-axisymmetric mode localized at the outer cylinder without stratification, and the instability of a mode propagating against the inner cylinder rotation with stratification. The combination of viscosity and stratification can also lead to a `collision' between (axisymmetric) Taylor vortex branches, causing the axisymmetric oscillatory state already observed in past experiments. Perhaps more surprising is the instability of a centrifugal-like helical mode beyond the Rayleigh line, caused by the joint effects of stratification and viscosity. The threshold $\mu=\eta^2$ seems to remain, however, an impassable instability limit for axisymmetric modes, regardless of stratification, viscosity, and even disturbance amplitude.

\end{abstract}



\maketitle


\section{Introduction}

There has been considerable recent interest in the effects of axial stratification on the Taylor--Couette problem following the discovery in 2001 \cite{MO01,YA01} that it leads to instabilities outside the centrifugally unstable region. This region is conventionally defined by Rayleigh's criterion \cite{RA17} as 
\begin{equation}
\mu \, < \, \eta^2\qquad 
 {\rm where} \quad \mu:=\Omega_o/\Omega_i\quad\text{and}\quad\eta:=r_i/r_o
\label{mu_eta}
\end{equation} 
are respectively the rotation and radius ratios between inner and outer cylinders (denoted with indices `$i$' and `$o$' respectively). Rayleigh derived his criterion for axisymmetric perturbations in the inviscid limit and only comparatively recently has it been extended to non-axisymmetric, inviscid perturbations by \cite{BI05}, albeit only in the limit of large axial wavenumbers. Using an inviscid, small-gap analysis, \cite{YA01,MO01} uncovered non-axisymmetric stratified instabilities that could develop when the inner cylinder rotates faster than the outer one, despite the radial decrease in angular momentum; the so-called quasi-Keplerian regime $\eta^2<\mu<1$. The new instability -- latter called the {\em stratorotational instability} or SRI in \cite{DU05a} -- was interpreted as a resonance between boundary-trapped inertia-gravity waves. Using the same asymptotic framework as \cite{BI05}, \cite{LE10} later showed that the SRI can become a radiative instability (RI) in the limit of an infinite gap ($\eta \to 0$) so that the outer boundary `goes to infinity'. The RI mechanism relies on a critical layer to extract energy from the base flow and radiate an evanescent wave radially outwards. More recently, \cite{PA13a} extended the instability range of stratified Taylor--Couette flow even further, reaching the striking conclusion that the flow is {\em always} unstable, except for the special case of solid-body rotation $\mu=1$. Importantly, \cite{PA13a} relaxed the small-gap assumption initially made by \cite{MO01,YA01} (by using large axial wavenumber asymptotics) and  uncovered the role played by a critical layer to achieve over-reflection between the two boundary-trapped waves causing SRI.

With the exception of \cite{WI74}, pre-2001 laboratory experiments on stratified Taylor--Couette flow were always carried out with a fixed outer cylinder \citep{WI74,BO95,BO96,CA99,CA00} (so $\mu=0\,<\,\eta^2$) and the relevance of the Rayleigh line was not questioned. The first experimental evidence of the SRI came in 2007 \cite{LE07} where non-axisymmetric instability was clearly observed in the centrifugally-stable regime. Significantly, \cite{LE07} explored a large range of rotation ratios and suggested a continuous connection between non-axisymmetric modes dominating on each side of the Rayleigh line. In contrast, \cite{LE10} claimed later that stratorotational instabilities (SRI/RI) are much weaker than centrifugal instabilities (CI) when $\mu<\eta^2$, implying that a) the SRI/RI and CI instabilities are distinct, and b) CI are always stronger. A distinction between SRI/RI and CI instabilities certainly exists in the inviscid limit (the optimal axial wavenumber is bounded for the SRI/RI \citep{PA13a,LE10} whereas it is not for CI \citep{BI05}) but this may not extend to the finite Reynolds numbers achieveable in experiments (consistent with \cite{LE07}). Certainly having this distinction a) simplifies the identification of which instability mechanism dominates at a given point in parameter space but is not guaranteed. Also, plausibly, CI might exist beyond the Rayleigh line in the presence of stratification (consistent with \cite{LE07}). To add to the uncertainty, statement b) seems inconsistent with the findings of \cite{PA13b} which showed that a RI could grow faster than a CI in the case of a stably-stratified Rankine vortex in a rotating frame.

The purpose of this paper is to shed some light on these seemingly contradictory statements by carrying out a stability analysis of stratified Taylor--Couette flow which bridges the gap between experimentally-relevant Reynolds numbers and large Reynolds numbers where inviscid analysis should hold in some fashion. The motivation for this study comes from an ongoing programme of experimental work \cite{L1,L2,L3} and the desire to be able to interpret the mechanistic origin of the instabilities observed there. The key questions to be addressed are as follows. 1) Are CI and SRI continuously `connected' (defined at the end of section \ref{explanation}) in parameter space or are they always distinct and thereby represent different instability mechanisms? 2) Can the SRI ever dominate CI in the centrifugally-unstable region $\mu < \eta^2$? And 3) can CI exist for $\mu > \eta^2$ and therefore beyond the Rayleigh line with stratification?

Viscous linear analyses already exist in the literature, starting with the contribution of \cite{TH66} based upon numerous simplifying assumptions. \cite{CA00} considered axisymmetric perturbations only (small-gap limit and finite gap but no density diffusion) while \cite{MO01,SH05,RU09} considered the general case albeit only close to the marginal stability curve. Here the focus is to consider the dispersion relation for linear viscous disturbances over several decades of the Reynolds number up from the marginal stability curve to at least $O(10^4)$ and sometimes $O(10^{10})$. A large range of rotation ratios is also considered in order to assess the relevance of the Rayleigh line $\mu=\eta^2$ at finite $Re$ and with stratification.  

The plan of the paper is as follows. In \S II, we briefly introduce the governing equations and numerical methods. In \S III, we present the results of our parametric stability analysis and discuss the effect of the different control parameters on the dominant mode. In \S IV, we assess the connection between CI and SRI at finite $Re$ by exploring the discontinuities of the optimal axial wavenumber in the $(\mu,Re)$-plane. By doing so, we reveal several connections between the different instabilities, and show the limits of Rayleigh's criterion in viscous stratified Taylor--Couette flow. We summarize our findings in \S V. 


\section{Problem formulation}

\subsection{Governing equations}

The geometry of an axially infinite container is characterised by the radius ratio $\eta$, while rotation and shear are jointly characterised by the rotation ratio $\mu$ (see (\ref{mu_eta})\,) and Reynolds number $Re$ which is defined as
\begin{equation}
Re:=\dfrac{r_i\Omega_i(r_o-r_i)}{\nu}
\end{equation}
with $\nu$ the kinematic viscosity. The purely azimuthal basic velocity profile $\mathbf{U}:=r\Omega(r)\bphi$ is of the form  
\begin{equation}
\Omega(r):=A+\dfrac{B}{r^2},
\label{base}
\end{equation}
with $A:=(\mu-\eta^2)/[\eta(1+\eta)]$ and $B:=\eta(1-\mu)/[(1+\eta)(1-\eta)^2]$.  The basic density profile $\bar{\rho}(z)$ is linear in the axial direction $z$ and characterised by a constant buoyancy frequency $N:=\sqrt{-(g/\rho_0)\mathrm{d}\bar{\rho}/\mathrm{d}z}$, based on gravity $g$ and a reference density $\rho_0$. Two nondimensional parameters characterise stratification: the Richardson and Schmidt numbers
\begin{equation}
Ri:=\dfrac{N^2}{\Omega_i^2}\quad\text{and}\quad Sc:=\dfrac{\nu}{\kappa},
\end{equation}
where $\kappa$ is the diffusivity of mass. For all the results presented in this paper, the Schmidt number was set to a value of 700 appropriate for salt-in-water experiments \cite{L1,L2,L3} with only a few exploratory calculations done with $Sc=7$ appropriate for heated-water experiments (see later in \S V). In the following, we take  $d:=r_o-r_i$ as the unit of length, $r_i\Omega_i$ as the unit of speed and $\rho_0 N^2(r_o-r_i)/g$ as the unit of density.

We consider the dynamics of infinitesimal perturbations of the velocity $\mathbf{u}'=u'\mathbf{\hat{r}}+v'\bphi+w'\mathbf{\hat{z}}$ in cylindrical coordinates, pressure $p'$ and density $\rho'$, around the linearly stratified basic flow (\ref{base}). Perturbations can be written in the form of normal modes
\begin{equation}
(u',v',w',p',\rho')=[u(r),v(r),w(r),p(r),\rho(r)]e^{\mathrm{i}(kz + m \theta-\omega t)}
\nonumber
\end{equation}
with complex frequency $\omega$, integer azimuthal wavenumber $m$ and real axial wavenumber $k$. In the Boussinesq approximation, the linearized incompressible Navier--Stokes, advection-diffusion and continuity equations read:
\begin{align}
\mathrm{i}su - 2 \Omega v  +\mathrm{d}_rp &= \dfrac{1}{Re}  \left( \nabla^2 u - \dfrac{u}{r^2} - \dfrac{2\mathrm{i}m}{r^2}v \right), \label{LS_1}\\
\mathrm{i}sv + Z u +\dfrac{\mathrm{i}m}{r}p &= \dfrac{1}{Re}\left(\nabla^2v - \dfrac{v}{r^2} + \dfrac{2\mathrm{i}m}{r^2} u\right), \label{LS_2}\\
\mathrm{i}sw +\mathrm{i}kp &=- Ri^* \,\rho + \dfrac{1}{Re} \nabla ^2 w,
\label{LS_3}\\
\mathrm{i}s\rho - w &= \dfrac{1}{Re \, Sc} \nabla^2 \rho, \label{LS_4}\\
0 &= \dfrac{1}{r} \mathrm{d}_r(ru) + \dfrac{\mathrm{i}m}{r}v + \mathrm{i}kw, \label{LS_5}
\end{align}
where $Ri^*:=Ri[(1-\eta)^2/\eta^2]$, $s := m \Omega - \omega$ is the Doppler-shifted frequency, $Z :=(1/r)\mathrm{d}(r^2 \Omega)/\mathrm{d}r$ is the axial vorticity of the basic flow and $\nabla^2 = \mathrm{d}^2_{rr} + (1/r)\mathrm{d}_r - (k^2 + m^2/r^2)$. The boundaries conditions are no slip ($u=v=w=0$) and no-flux $\mathrm{d}_r\rho=0$ at the walls. Symmetries are such that $\omega(k,m) = \omega(-k,m) = -\omega^{\star}(-k,-m)$, where $^{\star}$ denotes the complex conjugate. Therefore, we consider only positive $k$ and $m$ without loss of generality.
  
%
%
\subsection{Numerical methods}

The governing equations were discretized using Chebyshev collocation in the radial direction, leading to a generalized eigenvalue problem for $\omega$ and $(u,v,w,p,\rho)$. This problem was solved using multi-threaded LAPACK routines with OpenBLAS (\verb?http://www.openblas.net?). The number of Chebyshev polynomials used for each dependent variable was set to 140 for calculations below $Re=10^4$ and increased up to a maximum of 480 when required at higher $Re$ (at $(Re,\mu,\eta,Sc,m) = (10^4,0,0.417,700,1)$ and $Ri \in \{0.25,4,25\}$, doubling the resolution from 140 to 280 led to less than $1\%$ variation in the growth rate of the most unstable mode). The code was validated by reproducing figures 4 and 10 from \citep{PA13a}. For a given $m$, the most unstable mode was found by optimising the growth rate over $k$, using a Newton--Raphson method. Standard continuation methods were used to follow local maxima of the growth rate in parameter space.  
  

\section{Dominance diagrams \label{explanation}}

\begin{figure*}
\begin{center}
\includegraphics[width=0.8\textwidth]{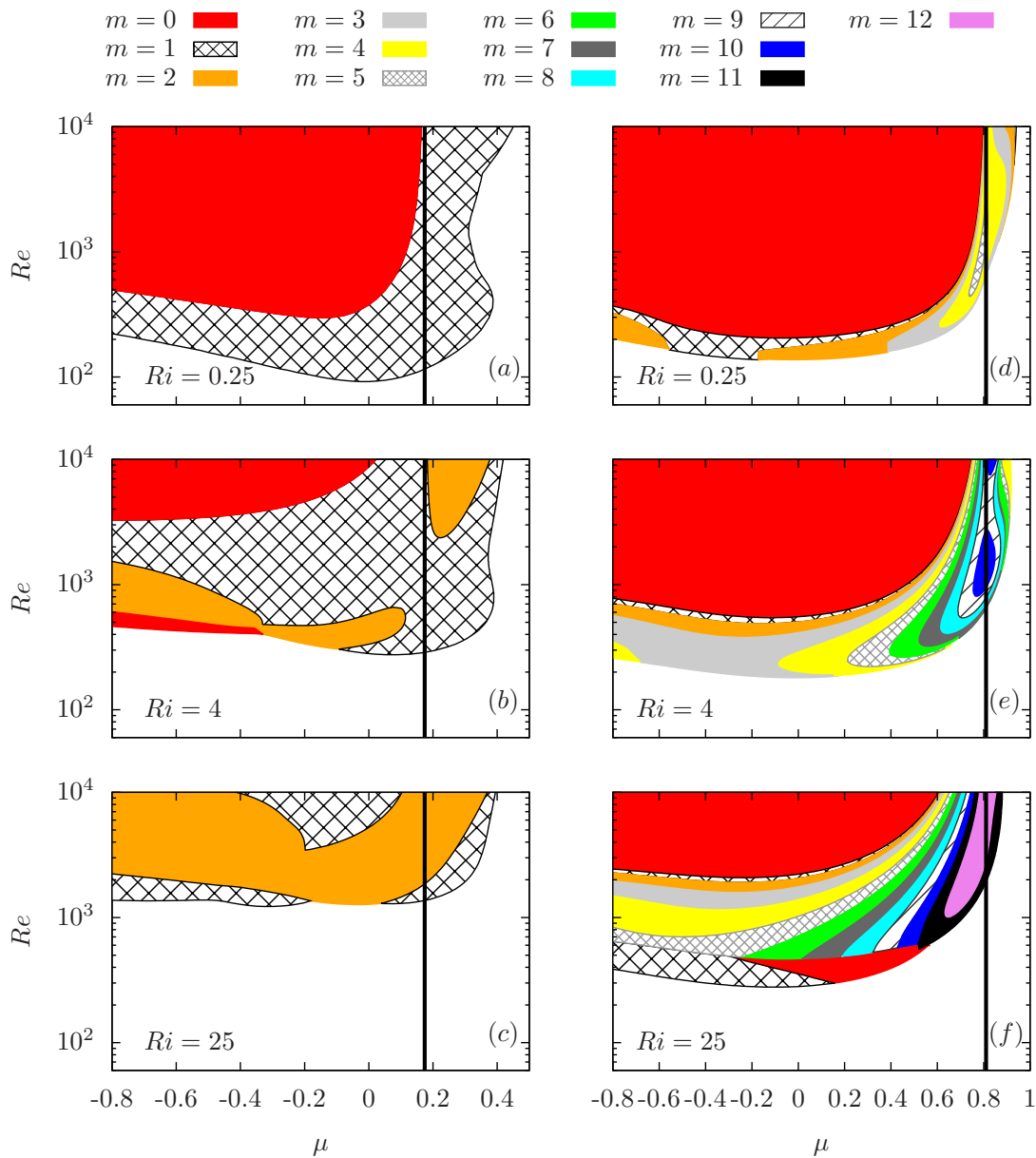}
\end{center}
\caption{Dominance diagrams showing the azimuthal mode number $m$ of the fastest growing mode (growth rate optimized over $k$) for $(a,b,c)$ $\eta=0.417$ and $(d,e,f)$ $\eta=0.9$ and $Ri=0.25,\,4$ and $25$ (indicated in the plot). The thick black line indicates the Rayleigh line.\label{fig:dominance_diagrams}}
\end{figure*}

We start by presenting the azimuthal mode number of the fastest growing mode in the $(\mu,Re)$-plane, for three values of $Ri=0.25,4 \, \& \, 25$ and two radius ratios: a `large' gap case $\eta=0.417$ (used in \cite{L1,L2,L3}) and a `small' gap case $\eta=0.9$. The dominant $m$ was obtained after optimization of the growth rate $\omega_\mathrm{i}:=\mathrm{Im}(\omega)$ over all possible sets of wavenumbers $(m,k)\in \mathbb{N}\times \mathbb{R^+}$. Results are given in figure \ref{fig:dominance_diagrams}, for a large range of $\mu$ and $Re\leq 10^4$. The vertical black line corresponds to the Rayleigh line $\mu=\eta^2$.

The first observation is the rise of the marginal stability curve to higher $Re$ indicating the stabilising effect of stratification for all $\mu$ and both $\eta$ considered, consistent with previous results in the literature. The second  common feature of all the plots is that axisymmetric ($m=0$) steady vortices (hereafter referred to as Taylor vortices) only dominate in regions removed from the marginal stability curve. The $m=0$ dominance regions near to the marginal curve are distinct since they have non-zero frequency. These oscillatory $m=0$ instabilities have already be found numerically by \cite{HU97,SH05}, and experimentally by \cite{CA99,CA00} but were not apparently seen by \cite{LE07} presumably because of their very restricted domain of dominance. Indeed, the critical instability is most often non-axisymmetric, with larger $m$ values becoming preferred as $Ri$ increases. In the small-gap case, the effect of the Rayleigh line is clearly visible: the dominant $m$ peaks to a maximum in its vicinity, and decreases on both sides. For the large-gap case though, only $m=0,1,2$ values dominate, and the Rayleigh line only seems to mark the limit of the dominant steady $m=0$ region.  

The dominance diagrams show the existence of disconnected patches and kinks in their boundaries (e.g. for $m=2$, $\eta=0.417$ and $Ri=4$). These features suggest that different instabilities corresponding to the same value of $m$ are competing for dominance. In order to assign an instability mechanism to each dominant mode, we track the loci of the discontinuities of $k_{\max}$ -- the axial wavenumber maximising the growth rate -- over $(\mu,Re)$-space for every fixed $m$ in the next section. A discontinuity in $k_{\max}$ indicates the coexistence of two global maxima in the growth rate curve $\omega_\mathrm{i}(k)$ (defined as the maximum growth rate at a {\em given} $k$): either side of this, the maxima switch dominance giving rise to the discontinuity in $k_{\max}$. If this discontinuity always separates the two competing instabilities in parameter space we refer to them as being {\em distinct} instabilities having different mechanisms. Conversely, if at some point the discontinuity terminates indicating that the local maxima have merged, we consider the two instabilities as being continuously {\em connected} in parameter space and therefore not distinct. (Formally, there is also the possibility that two distinct instabilities cross over momentarily having the same $\omega_\mathrm{i}$ at $k_{\max}$ but this would give rise to a discontinuity in $\omega_\mathrm{r}$ which is never seen in this study.)

\section{Exploring the discontinuities of the optimal axial wavenumber in the $(\mu,Re)$-plane}

\begin{figure*}
\begin{center}
\includegraphics[width=0.8\textwidth]{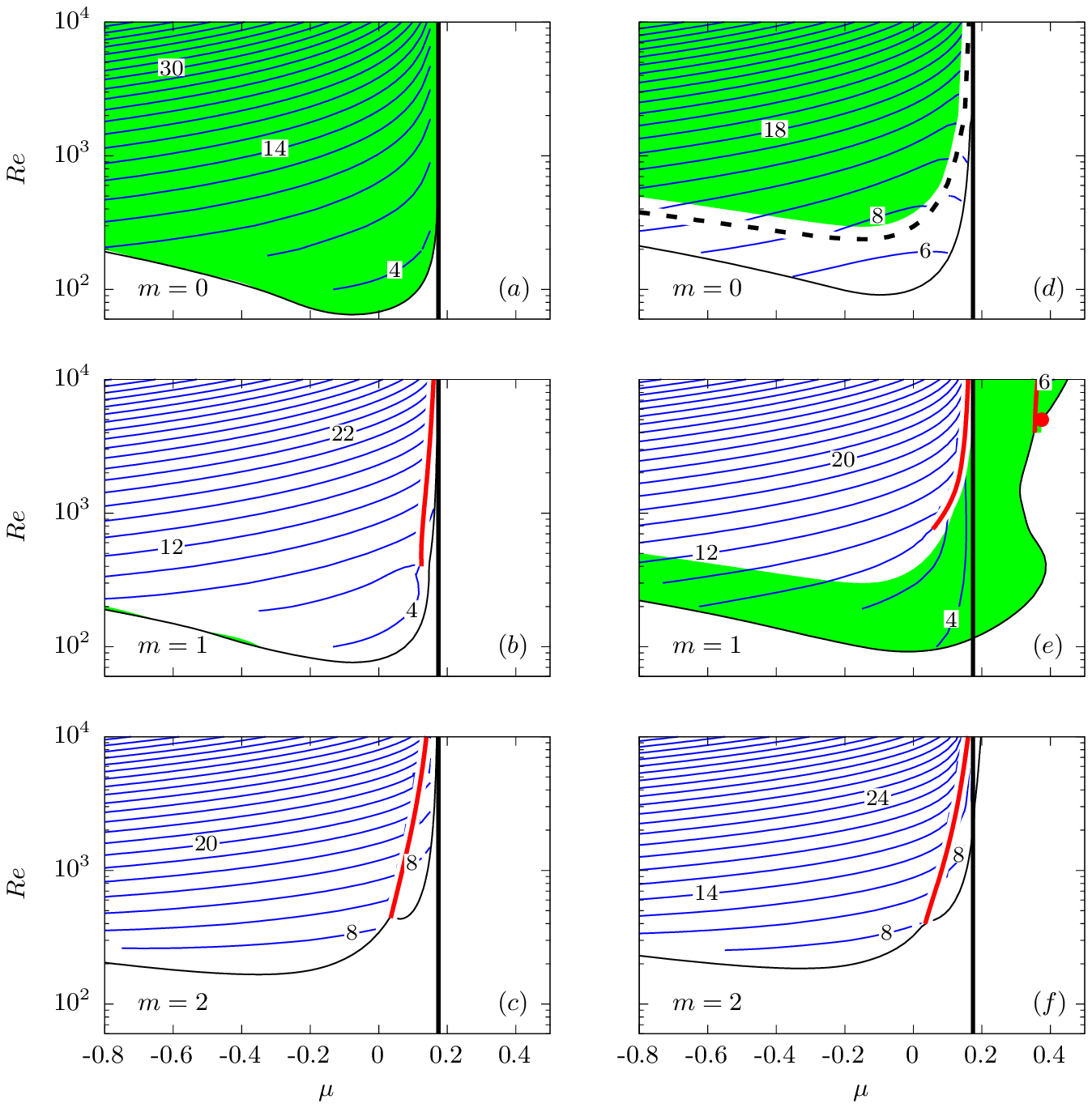}
\end{center}
\caption{Contours of optimal axial wavenumber $k_{\max}$ in the instability zone of modes $m=0$, $m=1$ and $m=2$ for $\eta=0.417$ and $(a,b,c)$ $Ri=0$, $(d,e,f)$ $Ri=0.25$. The spacing between contours is $\Delta k_{\max}=2$ and the maximum is always in the top left corner. The Rayleigh line $\mu=\eta^2$ is indicated by a thick black line. Discontinuities in $k_{\max}$ are indicated by thick red curves. Dashed curves for $m=0$ indicate a transition from an oscillatory (below) to a steady mode (above). The shaded regions indicate dominance of the given mode among all $m$ (note the dominance of $m=1$ in a very narrow range of parameter space for $Ri=0$ and  $\mu \in [-0.8,-0.4]$). Finally, the dot in the $(m,Ri)=(1,0.25)$ plot indicates the parameter values for the calculation of the eigenmode shown in figure \ref{fig:modes_strange}$(b)$.\label{fig:discont_k}}
\end{figure*}

\begin{figure*}
\begin{center}
\includegraphics[width=0.8\textwidth]{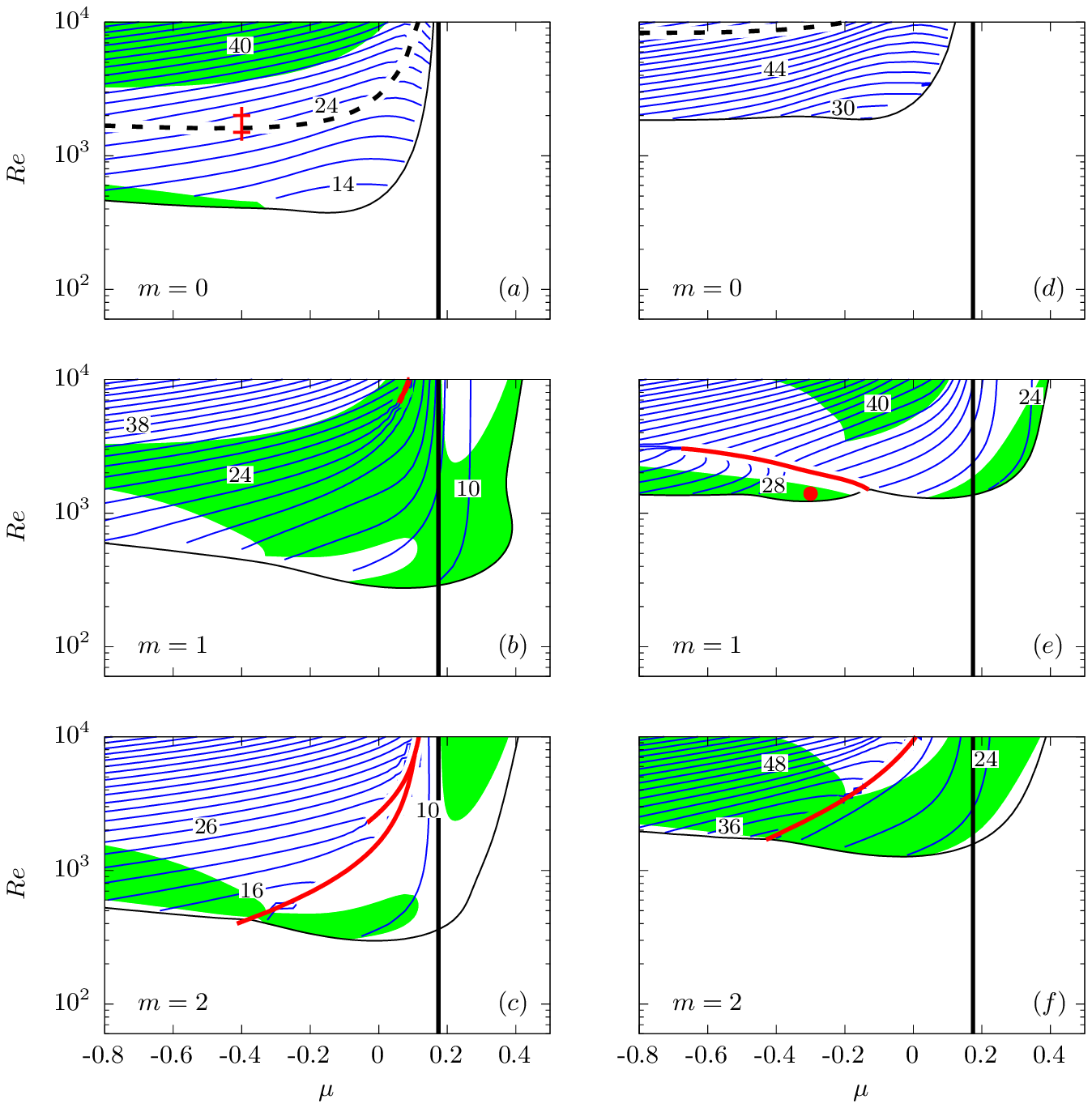}
\end{center}
\caption{Same caption as figure \ref{fig:discont_k} but  for $(a,b,c)$ $Ri=4$ and $(d,e,f)$ $Ri=25$. The dot in the $(m,Ri)=(1,25)$ plot  indicates the parameter values for the calculation of the eigenmode shown in figure \ref{fig:modes_strange}$(a)$. The red crosses in the $(m,Ri)=(0,4)$ plot indicate the parameter values used to generate figure \ref{fig:axi}.
\label{fig:discont_k2}}
\end{figure*}

Figures \ref{fig:discont_k} and \ref{fig:discont_k2} break down each dominance diagram into contributions from $m=0,1,2$ in the large-gap case. We plot the contours of $k_{\max}$, in order to identify the loci of the discontinuities for each $m$. The region where a given $m$ dominates overall is shaded. Figure \ref{fig:discont_k} compares the weakly stratified case ($Ri=0.25$) to the unstratified one, while figure \ref{fig:discont_k2} compares the moderate ($Ri=4$) and strong ($Ri=25$) stratifications.

\subsection{Oscillatory axisymmetric mode as a collision between Taylor vortex branches}

\begin{figure}
\begin{center}
\includegraphics[width=1.\columnwidth]{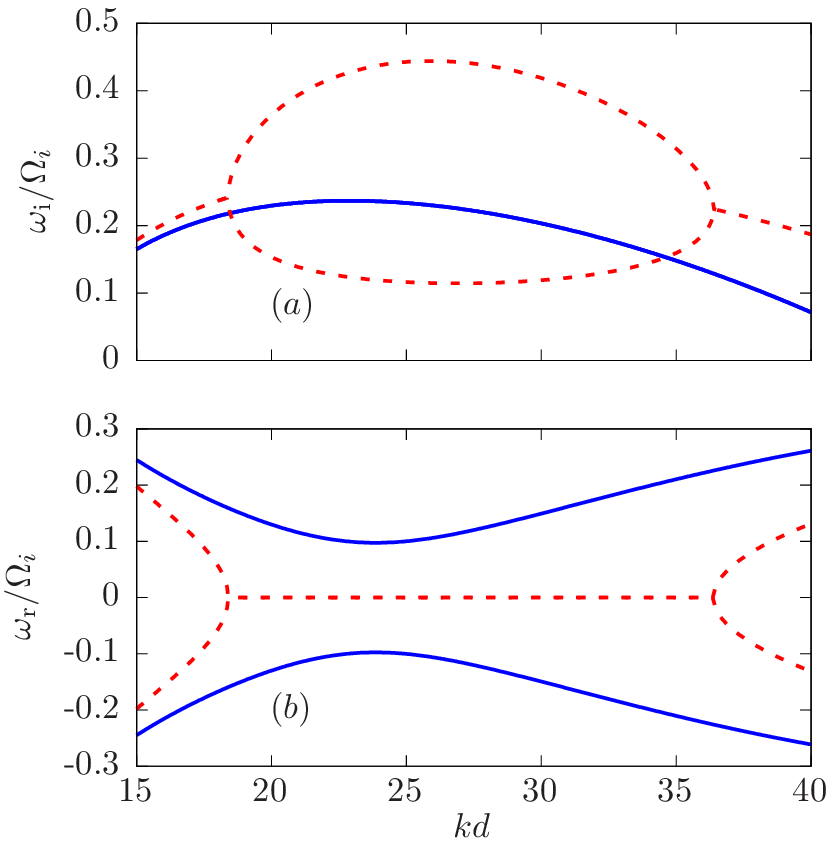}
\caption{$(a)$ Growth rate and $(b)$ frequency versus axial wavenumber $k$ of the two most unstable $m=0$ modes for $Ri=4$, $\mu=-0.4$ and $Re=1500$ (solid line) or $Re=2000$ (dashed line).\label{fig:axi}}
\end{center}
\end{figure}

For $m=0$, there is no discontinuity in $k_{\max}$ but a dashed curve marks the limit between oscillatory (below the line) and steady axisymmetric vortices (above). For all values of $Ri\neq 0$, the critical $m=0$ instability is always oscillatory, as already found by \cite{TH66,HU97} at large enough $Sc$, but becomes subdominant to steady vortices at large enough $Re$. To understand this transition better, we plot in figure \ref{fig:axi} the dependency of the frequencies $\omega_\mathrm{r}$ and growth rates $\omega_\mathrm{i}$ of the two dominant axisymmetric modes against $k$ for two values of $Re$: one below the dashed curve ($Re=1500$) and one above ($Re=2000$) (marked by red crosses in figure \ref{fig:discont_k2}). This figure shows that the oscillatory vortices are created as $Re$ decreases from $2000$ by the collision between two steady vortex branches. This is clearly a joint effect of stratification and viscosity, as the `bubble' in figure \ref{fig:axi}$(a)$ only appears if $Re$ is small enough and $Ri\neq 0$.

\subsection{Continuous connections between non-axisymmetric CI and SRI modes in the $(\mu,Re)$-plane}

\begin{figure}
\begin{center}
\includegraphics[width=1.\columnwidth]{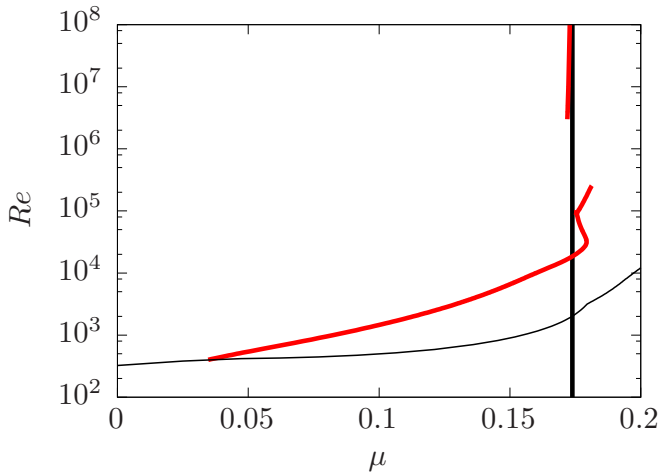}
\end{center}
\caption{Discontinuities of $k_{\max}$ in the $(\mu,Re)$-plane, for $\eta=0.417$, $m=2$, $Ri=0.25$ and $Re\in[10^2,10^8]$. As evident here, the discontinuity appearing in figure \ref{fig:discont_k} below $Re=10^4$ does not asymptote to the Rayleigh line as $Re\to \infty$. In fact, for $2\times 10^5\lesssim Re \lesssim 3\times10^6$ and $0.16<\mu<0.18$ there is no discontinuity at all in $k_{\max}$, which reveals, again, a continuous connection between CI and SRI instabilities at large, but finite $Re$. At even larger $Re$, a discontinuity reforms, permanently separating CI from SRI as $Re\to \infty$.}\label{fig:nodiscont}
\end{figure}

There are discontinuities in $k_{\max}$ for all non-axisymmetric modes, including when $Ri=0$, and these are indicated by thick red curves in figures \ref{fig:discont_k} and \ref{fig:discont_k2}. For $m=1$, the discontinuity near the Rayleigh line disappears within the unstable regions, at low enough $Re$, indicating that the instabilities on either side are smoothly connected. For $m=2$ however, the discontinuity always seems to separate the instability region into two distinct zones. But tracking the discontinuity further up in Reynolds number in figure \ref{fig:nodiscont} shows that it terminates just above $Re\approx 2 \times 10^5$ for $Ri=0.25$, indicating that the competing instabilities are again smoothly connected. 

At yet higher $Re$, a discontinuity reappears as it must do if the viscous analysis is to be consistent with inviscid predictions since there CI occurs at infinite $k$ \citep{BI05} whereas $k_{\max}$ remains finite for the SRI \cite{PA13a}. The value of $\mu$ where the growth rates of both instabilities are equal may be computed by suppressing the viscous term in the momentum equations (\ref{LS_1})--(\ref{LS_4}) for the SRI, and using the analytical expression given by \citep{BI05} for the CI in the inviscid limit (denoted with superscript $^\infty$): $\omega^\infty_\mathrm{i,CI}=\sqrt{-\Phi(r_i)}$ (independent of $m$), where $\Phi:=(1/r^3)\mathrm{d}(r^4\Omega^2)/\mathrm{d}r$ is the Rayleigh discriminant. We find that the switchover from SRI to CI dominant occurs at a value of $\mu$ which is very close to  but below $\eta^2$. Indeed, $\omega^\infty_{\mathrm{i,CI}}$ exactly vanishes on the Rayleigh line, whereas $\omega^\infty_{\mathrm{i,SRI}}$ does not. Therefore, there is a very narrow range of $\mu<\eta^2$ where non-axisymmetric SRI dominates CI in the inviscid limit. But this region is very small, which explains why \cite{LE10} claimed that CI dominates over SRI in the centrifugally unstable region. By a continuation argument, a discontinuity in $k_{\max}$ must be found at large but finite $Re$, which asymptotes the inviscid value of $\mu$ where $\omega^\infty_{\mathrm{i,SRI}}=\omega^\infty_{\mathrm{i,CI}}$. The discontinuities for $m=1$ in figures \ref{fig:discont_k} and \ref{fig:discont_k2} seem to directly approach this limit as $Re\to \infty$. For $m=2$ and $Ri=0.25$, the discontinuity forming at $Re\approx 3\times 10^6$ also asymptotes the inviscid limit at larger $Re$. This analysis of the discontinuities of $k_{\max}$ in the $(\mu,Re)$-plane establishes the continuous connection between CI and SRI instability mechanisms at finite $Re$ for $m=1$ and $m=2$.  

\subsection{A centrifugal instability mode localized at the outer cylinder in the unstratified case}

\begin{figure}
\begin{center}
\includegraphics[width=0.9\columnwidth]{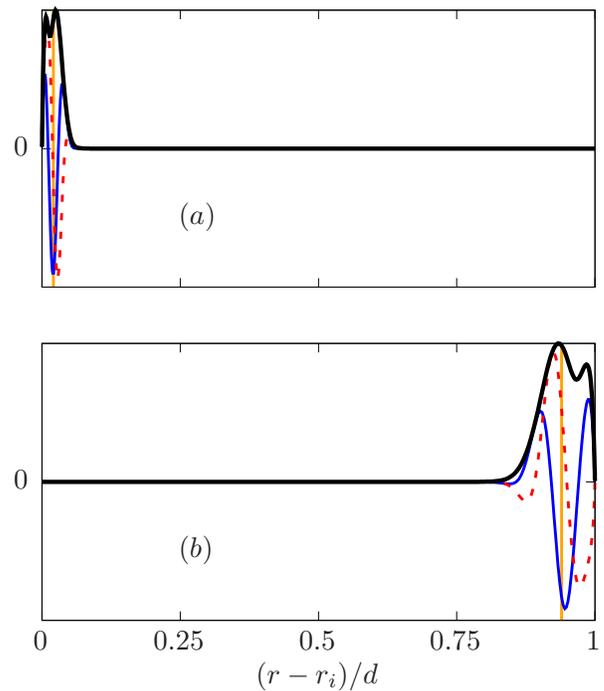}
\caption{Vertical velocity component of the two leading $m=1$ eigenmodes at the discontinuity in $k_{\max}$ for $Re=10^6$ and $Ri=0$: $(a)$ left side of the discontinuity, $(b)$ right side. Thin solid line: real part $w_\mathrm{r}$; dashed solid line: imaginary part $w_\mathrm{i}$; thick solid line: $|w|$. Points where $\Omega(r_t)-\omega=0$ are indicated with a solid line at $r=\mathrm{Re}(r_t)$. These correspond to turning points in the WKB theories of RI \citep{PA13a} and SRI \citep{LE10}, but do not play any particular role in the asymptotic description of CI \citep{BI05}. Here we show that these points indicate the position where the modes are localized at large $Re$. \label{fig:modes_1e6_Ri0}}
\end{center}
\end{figure}

A simple way to attempt to differentiate between CI and SRI is to suppress stratification: if the flow is stabilized, the instability was a SRI, otherwise it was a CI. This motivated us to compute the dispersion relation of modes $m=0,1,2$ in the unstratified case as well. As shown in figure \ref{fig:discont_k}, there is no longer an instability beyond the Rayleigh line when $Ri=0$, but surprisingly, we still observe a discontinuity in $k_{\max}$ asymptoting $\mu=\eta^2$ at large $Re$ for $m=1$ and $2$. This result was unexpected, as we previously associated such discontinuity to a CI and SRI instability competing in the inviscid limit. However, since SRI is caused by a resonance between boundary-trapped inertia-gravity waves \citep{MO01,YA01}, the mode dominating on the right of the discontinuity cannot be SRI when $Ri=0$. This suggests that this previously-unreported CI branch on the right of the discontinuity for $Ri=0$ is connected to a SRI as $Ri$ increases from 0. This new CI mode dominating to the right of the discontinuity is localized exclusively at the outer cylinder, as can be seen in figure \ref{fig:modes_1e6_Ri0}$(b)$ at $Re=10^6$, whereas the one to the left of the discontinuity is localized at the inner cylinder (figure \ref{fig:modes_1e6_Ri0}$(a)$). These wall modes are reminiscent of the two families of neutral branches which create SRI in the inviscid limit \citep{PA13a}. In that limit, the outer-wall mode can only become unstable by coupling with the inner one, under the effect of stratification, but here we find that it may become unstable alone with the help of viscosity. Since the discontinuity in $k_{\max}$ asymptotes to the Rayleigh line as $Re \to \infty$, this mode never dominates the inner-wall one in the inviscid limit: there is therefore no contradiction with the theoretical analysis of \cite{BI05}.

\subsection{Radiative instability mode connecting CI to SRI in finite gap}

\begin{figure}
\begin{center}
\includegraphics[width=1.\columnwidth]{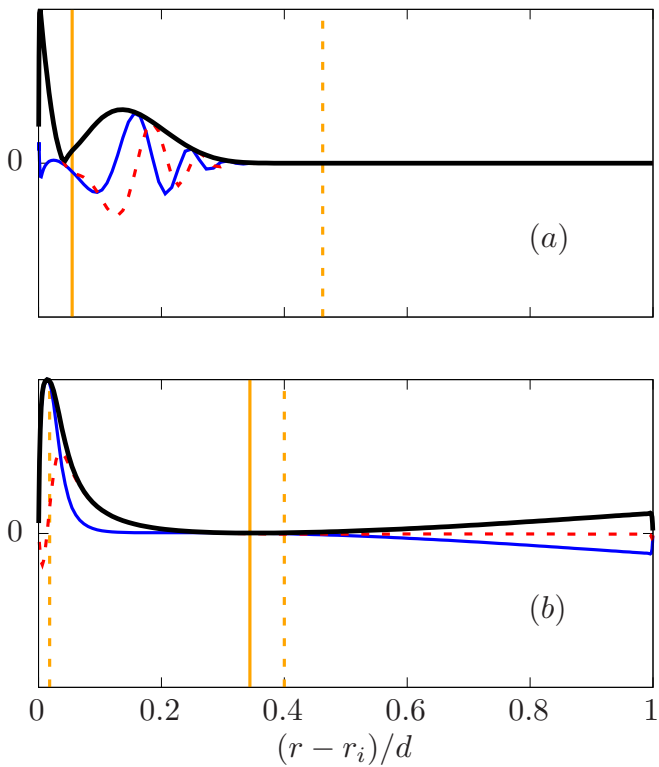}
\caption{Same caption as \ref{fig:modes_1e6_Ri0}, but for the leftmost discontinuity (near the Rayleigh line) in figure \ref{fig:discont_k} for $(m,Ri)=(1,0.25)$ and  $Re=10^6$. Here we also indicate the critical points $\Omega(r_c^\pm)-\omega=\pm N$ with dashed lines at $r=\mathrm{Re}(r_c\pm)$ ($r_c^-$ is outside the domain in $(a)$).\label{fig:modes_1e6}}
\end{center}
\end{figure}

In figure \ref{fig:modes_1e6}, we compare the structure of the two modes on the discontinuity close to the Rayleigh line for $m=1$, $Re=10^6$ and $Ri=0.25$. The instability dominating to the left (decreasing $\mu$) of the discontinuity must tend to a CI as $Re\to\infty$, while the instability to the right must tend to a SRI. 
As way of confirmation, the right branch has a structure which is reminiscent of inviscid SRI: the mode is localized at the walls and has a critical layer (see figure \ref{fig:modes_1e6}$(b)$), as described  by \cite{PA13a}. However, since $k_{\max}$ remains small, the WKB framework of \cite{PA13a} does not obviously apply so there are not the oscillatory regions described by these authors. 

The structure of the left branch, however, resembles the radiative instability mode described by \cite{LE10} in the limit where the gap and Reynolds numbers become infinite, while $\mu \to \eta^2$ (and $kd \gg 1$). There is a critical layer and an oscillatory region of radially decaying amplitude in figure \ref{fig:modes_1e6}$(a)$, similar to figure 3 in \cite{LE10}. This region is bounded to the right by the critical point $\Omega(r_c^+)-\omega=N$, as in these authors' theory for weak stratification. This critical point effectively isolates the radiated wave from the outer cylinder, which may explain why we were able to find a RI mode in our finite gap geometry, whereas \cite{LE10} only refer to this instability in the limit of infinite gap. In the present case, the RI seems to mediate the continuous morphing from the SRI to a CI at finite $Re$. 

\subsection{A helical mode propagating against the inner cylinder rotation}

\begin{figure}
\begin{center}
\includegraphics[width=1.\columnwidth]{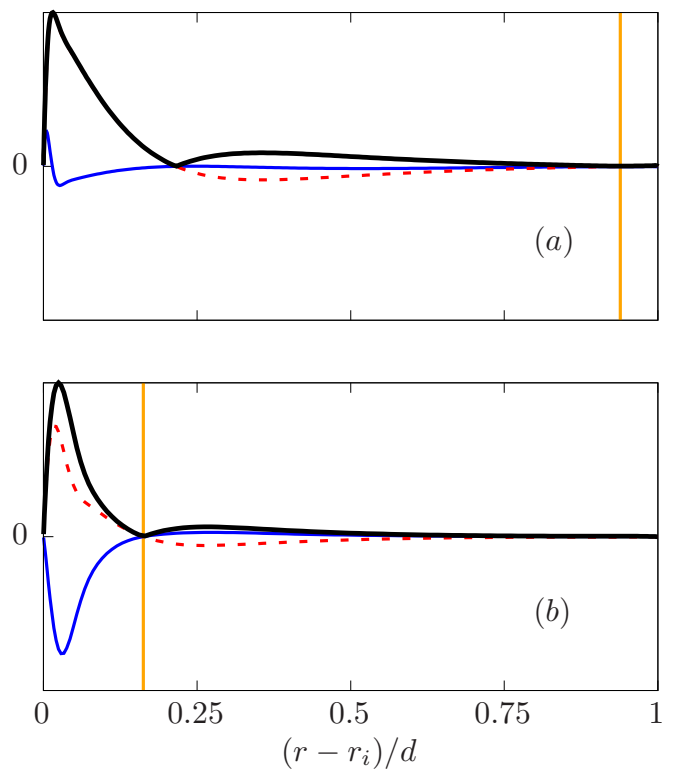}
\caption{Vertical velocity component of the leading eigenmodes at parameter values indicated by solid dots in figures \ref{fig:discont_k} and \ref{fig:discont_k2}.$(a)$ Mode with negative azimuthal velocity at $\mu=-0.3$, $Re=1400$, $m=1$, $Ri=25$. $(b)$ Centrifugal-type mode destabilized by viscosity beyond the Rayleigh line, at $\mu=0.375$, $Re=5000$, $m=1$, $Ri=0.25$. Solid and dashed curves: same caption as in figures \ref{fig:modes_1e6_Ri0} and \ref{fig:modes_1e6}. Turning points $r_t$, as defined in figure \ref{fig:modes_1e6_Ri0} are shown, but critical points $r_c^\pm$, as defined in figure \ref{fig:modes_1e6}, are outside the domain in both $(a)$ and $(b)$.\label{fig:modes_strange}}
\end{center}
\end{figure}

Finally, in the last two subsections we investigate the nature of the modes in regions of the $(\mu,Re)$-plane created by unexpected discontinuities of $k_\text{max}$, i.e. discontinuities that do not separate CI from SRI in the inviscid limit. First we discuss the critical $m=1$ mode at large $Ri=25$ in the counterrotating regime $ \mu <0$ (see the solid dot in figure \ref{fig:discont_k2}). Surprisingly, the azimuthal velocity $\omega_\mathrm{r}/m$ associated with that mode is negative, whereas it is always positive for other dominant modes. Such peculiar behaviour has not been reported before. Inspecting the structure of this mode in figure \ref{fig:modes_strange}$(a)$ indicates that it is not a SRI, as the amplitude of $|w|$ is negligible near the outer wall. The turning point $r_t$ such that $\Omega(r_t)-\omega=0$ does not seem to coincide with any particular feature of the mode structure, confirming that asymptotic theories of CI/RI/SRI are of little help at this low $Re=1400$. Because of the apparent absence of a critical layer, we conclude that the mode is best classed as a CI.

\subsection{A centrifugal-type mode beyond the Rayleigh line destabilized by viscosity}

\begin{figure}
\begin{flushright}
\includegraphics[width=1.\columnwidth]{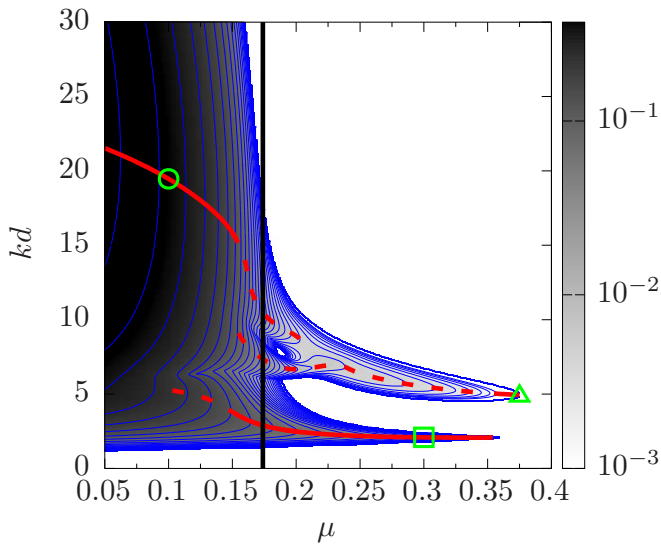}
\end{flushright}
\caption{Positive growth rate contours (with $\Delta \log_{10}\omega_\mathrm{i}/\Omega_i=0.1$) in the $(\mu,k)$-plane for $Re=5000$, $Ri=0.25$ and $m=1$. The Rayleigh line $\mu=\eta^2$ is indicated with a solid vertical line. Values of $k$ corresponding to local maxima of the growth rate at the given $\mu$ are indicated with thick dashed curves, becoming solid when the maximum is global.\label{fig:nice}}
\end{figure}

Finally, we discuss the discontinuity located at the right of the Rayleigh line in figure \ref{fig:discont_k} for $(m,Ri)=(1,0.25)$. Since the dominant mode is `beyond' the Rayleigh line and well-separated from the CI region by two discontinuities in $k_\text{max}$, it is tempting to call this mode SRI. However, looking at the mode in figure \ref{fig:modes_strange}$(b)$ reveals a structure which is highly reminiscent of the CI found in the previous section. The only noticeable difference is the presence of a turning point (as defined in the previous section) exactly where $|w|=0$, suggesting the mode may extract its energy from the base flow at the critical layer, as in the RI. In order to determine whether the mode is centrifugal or radiative, we plot isocontours of the growth rate in the $(\mu,k)$-plane in figure \ref{fig:nice}. The local maxima of the growth rate in that plane are indicated with dashed red curves, becoming solid when the maximum is global. The plot shows that the dominant (only) instability at $\mu=0.375$  is created by the merging of two local maxima present in the `centrifugally-unstable region' $\mu < \eta^2$. This may explain why the mode is both reminiscent of CI and RI.

\begin{figure}
\begin{center}
\quad
\includegraphics[width=1.\columnwidth]{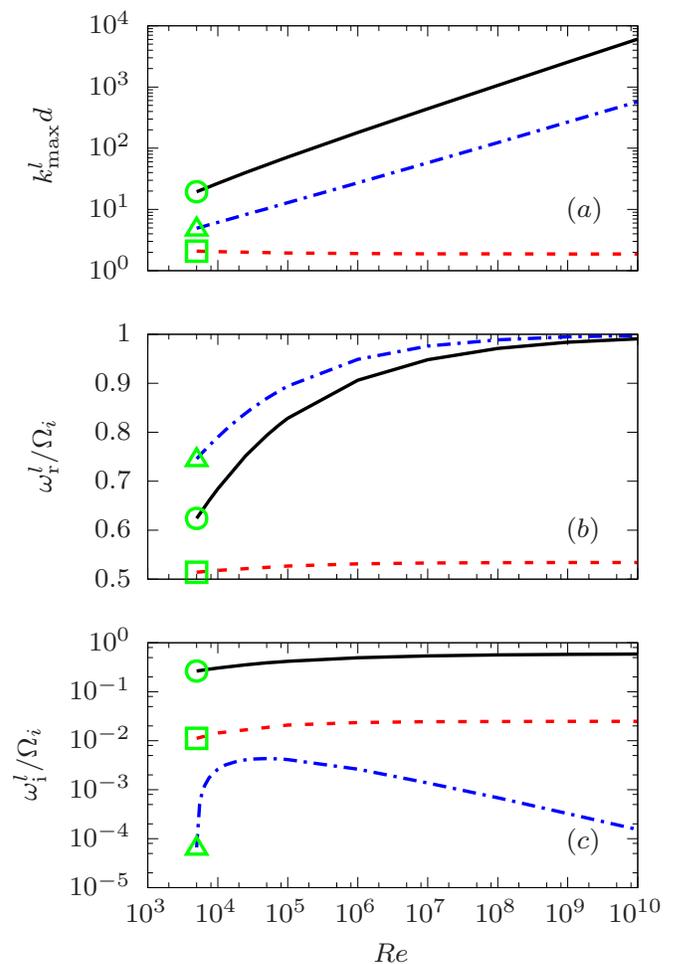}
\end{center}
\caption{Evolution of $(a)$ the local optimal axial wavenumber $k^l_{\max}$ and its associated $(b)$ frequency $\omega^l_\mathrm{r}$ and $(c)$ growth rate $\omega^l_\mathrm{i}$ versus Reynolds number for fixed $\mu=0.1$ (solid line), $\mu=0.3$ (dashed line) and $\mu=0.375$ (dash-dotted line); $Ri=0.25$ and $m=1$. The wavenumber $k^l_{\max}$ maximizes the growth rate globally when $Re=5000$ (cf. figure \ref{fig:nice}).}\label{fig:followRe}
\end{figure}

Pursuing this further, the three distinct branches, respectively dominating at $\mu=0.1,0.3 \, \& \, 0.375$ are followed  as $Re\to\infty$ in figure \ref{fig:followRe}. Each branch is a global maximum of $\omega_\mathrm{i}(k)$ at $Re=5000$ and its respective value of $\mu$, but only a local maximum at larger $Re$, hence the superscript $^l$ to denote the associated $k_{\max}$, $\omega_\mathrm{r}$ and $\omega_\mathrm{i}$. In figure \ref{fig:followRe}, we observe very similar trends for the evolution of $k^l_\text{max}$ and $\omega^l_\text{r}$ between the unknown mode at $\mu=0.375$ and the CI at $\mu=0.1<\eta^2$. Indeed, in both cases, $k^l_\text{max} \to \infty$ and $\omega^l_\text{r}/\Omega_i\to 1$. Both trends were predicted analytically for ($m=1$) inviscid CI by \cite{BI05}, and that  $k_{\max}$ remains finite and $\omega_\mathrm{r}/\Omega_i \,<\, 1$ for the RI \cite{LE10}. This suggests that the mode dominating at $(\mu,Re)=(0.375,5000)$ connects with CI at large $Re$, despite being outside the so-called centrifugally unstable region. This observation does not contradict the theory in  \cite{BI05} as the growth rate of this mode tends to zero in the inviscid limit. The conclusion is therefore that this mode is a mixture between a CI and a RI mode, destabilised beyond the Rayleigh line by the joint effects of stratification and viscosity. 

Our numerical results suggest that this intrusion of CI-type mode features beyond the Rayleigh line is only possible for non-axisymmetric modes. That {\em inviscid} axisymmetric modes still can't exist beyond the Rayleigh line with stable stratification follows from a simple extension of Rayleigh's criterion (see appendix A). A stronger result can be proved that no {\em finite-amplitude} axisymmetric state beyond the simple base flow can exist past the Rayleigh line for a viscous, stratified fluid in the thin-gap limit $\eta \to 1$ (see appendix B). But whether a general proof exists for finite curvature, viscosity and stable stratification is unclear.

\section{Conclusions}

In this paper, we have performed a thorough linear stability analysis of (axially) stably-stratified viscous Taylor--Couette flow motivated by ongoing experiments \cite{L1,L2,L3}. We considered both a large-gap $\eta=0.417$ and a small-gap $\eta=0.9$ configuration and varied the rotation ratio $\mu$, the Reynolds number $Re$ and the Richardson number $Ri$. This paper bridges the gap between the numerous inviscid analyses of the centrifugal, stratorotational and radiative instabilities (respectively CI, SRI and RI), and the computations of instability thresholds in the viscous case, by addressing the question of the dominant instability mechanism at finite $Re$ beyond the marginal stability line. 

As in previous studies, we found that the first bifurcation as $Re$ increases above the marginal curve is always oscillatory, and usually non-axisymmetric. We showed that the oscillatory $m=0$ mode is created by a collision between steady Taylor vortex branches. This is an effect of viscosity and stratification (and large $Sc$ too, according to \cite{TH66,HU97}), as this collision disappears at large $Re$ or for $Ri \neq 0$.

By exploring the discontinuities in the optimal axial wavenumber $k_{\max}$, we were able to establish a continuous connection between CI and SRI instabilities in the $(\mu,Re)$-plane. Indeed the optimal wavenumber $k_{\max}$ diverges for CI in the inviscid limit, whereas it remains finite for the SRI: as a consequence, there exists a value of $\mu\lesssim \eta^2$ where the growth rate of both instabilities are equal. This leads to a discontinuity in $k_{\max}$ at $Re\to\infty$ which can be tracked down to finite $Re$, where it eventually disappears as the two local maxima of $\omega_\mathrm{i}(k)$ merge into one. The coalescence occurs within the instability region, so it is impossible to distinguish CI from SRI at onset. This explains why \cite{LE07} observed a smooth evolution of non-axisymmetric patterns from the quasi-Keplerian to the centrifugally-unstable region in their experiment, whereas \cite{LE10} considered the two instabilities distinct in the inviscid limit. As argued by \cite{LE10}, CI is indeed stronger than SRI when the two instabilities compete in the inviscid limit, but the discontinuity in $k_{\max}$ bends back towards low values of $\mu$ as $Re$ diminishes, making SRI dominant over a large portion of the Rayleigh-unstable domain, until the two instabilities can no longer be distinguished. We also established a connection between CI and RI induced by viscosity, without taking the limit of infinite gap as in \cite{LE10}. 

In the unstratified case, we did not expect to find similar discontinuities in $k_\text{max}$ for $m\neq0$, as SRI is a resonance between boundary-trapped inertia-gravity waves. However, we did find two different branches competing in the vicinity of the Rayleigh line for a given $m\neq 0$ when $Ri=0$. Each of the competing branches corresponds to a wall-mode, localized either at the inner or the outer boundary. These are reminiscent of the two families of branches giving birth to SRI as they interact when $Ri\neq 0$ \cite{PA13a}. But here the outer-wall mode becomes unstable under the effect of viscosity, not because of a coupling with the inner-wall mode. This outer-wall CI mode, continuously connected to SRI as $Ri\neq 0$ was previously unknown because it is always subdominant, yet it is destabilized at finite $Re$ and large-gap widths.

We have also found a dominant $m=1$ mode propagating against the rotation of the inner cylinder in the counterrotating strongly stratified case. This behaviour has not been reported before, for either one of the three instabilities studied in this paper, since they all require $0\leq\omega_\mathrm{r}/m<\Omega_i$  to be able to extract energy from the base flow when $Re \to \infty$. This is another surprising effect of viscosity on the instability mechanism, since this mode is only dominant at low enough $Re$.

Finally, we investigated the nature of a $m=1$ mode dominating beyond the Rayleigh-line in the weakly stratified case. This mode seems to have a critical layer, but since its maximum amplitude peaks at the inner cylinder only, it is reminiscent of a radiative mode. However, it shares features of CI, rather than RI modes as $Re \to \infty$. In particular, the optimal wavenumber diverges while the azimuthal phase speed tends to the inner cylinder angular velocity: two properties of CI. This, however, does not violate the generalized Rayleigh's criterion of \cite{BI05} as the growth rate asymptotes zero in the inviscid limit. At finite $Re$, this mode seems connected to both CI and RI in the $(\mu,k)$-plane, therefore, we conclude that it is a mixture of CI and RI, destabilized beyond the Rayleigh line by the coupled effects of viscosity and stratification. 
This reiterates that viscosity has a more complex impact on the stability properties beyond just expected stabilization.

For $\eta=0.9$, we did not systematically investigate the discontinuities of $k_{\max}$ in the $(\mu,Re)$-plane. None were found for both $m=1$ and $m=2$ at $Ri=0.25$ in the range $Re<10^4$ but since a discontinuity must be present near the Rayleigh line in the inviscid limit, we conclude that these discontinuities form at larger $Re$ as $\eta \to 1$. This makes the distinction between SRI-type and CI-type instabilities even more problematic, and indicates that the effect of viscosity is heightened by reducing the gap size. We also produced dominance diagrams at a lower value of the Schmidt number $Sc=7$, in the large-gap case, for $Ri=4 \, \& \,25$ which are qualitatively similar to $Sc=700$ and so not included. The $Sc$ number effects are expected to occur at even lower values of $Sc$, according to \cite{TH66,HU97}.

We conclude by assessing the relevance of the Rayleigh line: even though non-axisymmetric centrifugal-type modes seem to be able to grow beyond $\mu=\eta^2$ with the help of viscosity and stratification, this limit appears to remain impassable to axisymmetric ones (whether steady or oscillatory). We were able to prove this result in two distinct limits: inviscid linear disturbances in finite gap and viscous finite-amplitude disturbances in thin gap. Whether a more general result suggested by our numerical results (and previous studies) can be proven remains an interesting question.

\section*{Acknowledgements}
This work has been supported by the EPSRC (C. L.), under grant EP/K034529/1, and \'Ecole Normale Supérieure de Paris (F. N.). We thank St\'ephane Le Diz\`es for stimulating discussions.

\appendix

\section{Rayleigh's Criterion for Stably-Stratified Inviscid Flow}

The Euler equations linearised around the basic flow $\bU:=r \Omega(r) \bphi$ (equations (\ref{LS_1}-\ref{LS_5}) with $Re \rightarrow \infty$) for an axisymmetric ($m=0$) incompressible disturbance can be reduced down to a 2\textsuperscript{nd} order differential equation for $u$, the radial perturbation velocity,
\beq
\dfrac{\mathrm{d}}{\mathrm{d}r} \dfrac{1}{r} \dfrac{\mathrm{d}(ru)}{\mathrm{d}r}=\dfrac{k^2 (s^2-\Phi(r))}{s^2-Ri^*} u \quad
{\rm where} \quad
\Phi:=\dfrac{1}{r^3} \dfrac{\mathrm{d} (r^2 \Omega)^2}{\mathrm{d}r}
\eeq 
is the Rayleigh discriminant (\cite{DR81}, p. 69).
Multiplying by $ru^*$ (where $u^*$ is the complex conjugate of $u$) and integrating from $r_i$ to $r_o$ gives
\begin{align}
-\int^{r_o}_{r_i} \dfrac{1}{r}\biggl| \dfrac{\mathrm{d}(ru)}{\mathrm{d}r}\biggr|^2 \,dr
&= 
\dfrac{k^2 s^2}{s^2-Ri^*} \int^{r_o}_{r_i} r\left|u\right|^2 \,dr  \nonumber\\
& \!\!\!\!\!-\dfrac{k^2}{s^2-Ri*} \int^{r_o}_{r_i} r\Phi(r)\left|u\right|^2 \,dr 
\end{align}
since $u(r_i)=u(r_o)=0$. Rearranging
\beq
s^2= \dfrac{Ri^* \int^{r_o}_{r_i} \dfrac{1}{r}\biggl|\dfrac{\mathrm{d}(ru)}{\mathrm{d}r}\biggr|^2 \,dr+k^2\int^{r_o}_{r_i} r\Phi(r)\left|u\right|^2 \,dr}
               {\int^{r_o}_{r_i} \dfrac{1}{r}\left| \dfrac{\mathrm{d}(ru)}{\mathrm{d}r}\right|^2 \,dr+k^2 \int^{r_o}_{r_i} r\left|u\right|^2 \,dr}
               \label{criterion}
\eeq
so providing  $\Phi(r)>0$ for all $r \in (r_i,r_o)$, i.e. the basic flow is Rayleigh-stable, then all of the integrals are positive definite which implies $s_\mathrm{i}=0$ and stability for all $k$. 

\section{Uniqueness of axisymmetric states beyond the Rayleigh line in the thin gap limit}

Here we prove that the only {\em streamwise-independent} state that can exist in rotating, 
stably-stratified plane Couette flow beyond the Rayleigh line is one of simple shear implying that no other axisymmetric state beyond the base state can exist beyond the Rayleigh line in thin-gap stratified Taylor--Couette flow. The proof is a straightforward extension of the unstratified result presented by \cite{HU72} to include stratification. In a rotating frame $\bOmega=\Omega \bhz$ where the shearing boundaries are at $y=\pm 1$ and gravity $\mathbf{g}:=-g\bhz$,
there is the  simple shear solution $\mathbf{U}=y \bhx$, $P=-\Omega y^2+\half Ri\, z^2$ and $\overline{\rho}=-z$ (stable stratification). Rayleigh's criterion in this context is that centrifugal instability is only possible for $\Omega < \half$ (e.g. see \cite{RI07}). The governing equations for disturbances away from this steady state, $(\bu:=\mathbf{u}_\text{tot}-\mathbf{U},p:=p_\text{tot}-P,\rho:=\rho_\text{tot}-\overline{\rho})$ are
\begin{align}
\partial_t \mathbf{ u} +2 \Omega \bhz \times \mathbf{u}+& y \partial_x \mathbf{u}+  v \bhx+\mathbf{u} 
\cdot \nabla \mathbf{u}  = \hspace{1cm}\nonumber \\
& -\nabla p - Ri \, \rho  \, \mathbf{\hat{z}}
   + \dfrac{1}{Re} \nabla^2 \mathbf{u}, \label{b1}\\
\partial_t \rho +y \partial_x \rho-w+ & \mathbf{u}  \cdot \nabla \rho =  \dfrac{1}{Re \, Sc} \nabla^2 \rho, \label{b2}\\
&\nabla \cdot \mathbf{u}  = 0. \label{b3}
\end{align}
Defining $\langle (\, \cdot\,) \rangle:=\dfrac{1}{2L} \int^L_0 \int^1_{-1} (\, \cdot\,) {\rm dydz}$ and $\mathbf{u}=u \bhx+v \bhy+w \bhz$, 
then for streamwise-independent velocity and density fields
taking $\langle u \bhx \cdot (\ref{b1})\rangle$, $\langle  (v \bhy +w \bhz) \cdot (\ref{b1}) \rangle$ and $\langle \rho (\ref{b2})\rangle$ leads to the `energy' integrals,
\begin{align}
\langle \half u^2      \rangle_t &= (2 \Omega-1) \langle u v \rangle-\dfrac{1}{Re} \langle |\bnab u|^2 \rangle,  \label{E_u}\\
\langle \half (v^2+w^2) \rangle_t &= -2\Omega \langle u v\rangle 
                                     -Ri \langle \rho w \rangle \nonumber \\
                                   & \hspace{1cm}  -\dfrac{1}{Re} \langle |\bnab v|^2 +|\bnab w|^2 \rangle, \label{E_vw}\\
\langle \half \rho^2   \rangle_t &=  \langle \rho w \rangle -\dfrac{1}{Re\,Sc} \langle |\bnab \rho|^2 \rangle \label{E_rho} 
\end{align}
where, periodicity across $z \in [0,L]$ and either non-slip or stress-free velocity fields together with either Dirichlet ($\rho=0$) or Neumann conditions ($\partial_n \rho=0$) for the density on $y=\pm 1$ kill  all boundary terms which arise.
Importantly, all the cubic nonlinear terms drop in these equations and so the kinetic energy in the $u$ field can be treated separately from that in $v$ and $w$. As a result,  generalised energy  and dissipation functionals can be constructed as follows
\begin{align}
E_\lambda:=&\half \langle \lambda^2 u^2    +v^2 +w^2 + Ri \rho^2 \rangle, \\
D_\lambda:=&      \langle  \lambda^2|\bnab u|^2+| \bnab v|^2+|\bnab w|^2+ \dfrac{Ri}{Sc}|\bnab \rho|^2 \rangle.
\end{align}
Then $\lambda^2 (\ref{E_u})+(\ref{E_vw})+Ri (\ref{E_rho})$ gives
\beq
\dfrac{\mathrm{d} E_{\lambda}}{\mathrm{d}t}= D_\lambda \biggl\{ \dfrac{[\,2 \Omega(\lambda^2-1)-\lambda^2\,]\langle u v \rangle}{D_\lambda}-\dfrac{1}{Re}  \biggr\}
\label{E_eqn}
\eeq
and monotonic decay of the disturbance energy is ensured if
\beq
\dfrac{1}{Re} \,>\, \text{max}_{u,v,w,\rho} 
\, \, \dfrac{[\,2 \Omega(\lambda^2-1)-\lambda^2\,]\langle \,u v \,\rangle}
     {\langle \, \lambda^2|\bnab u|^2+| \bnab v|^2+|\bnab w|^2 + \dfrac{Ri}{Sc}|\bnab \rho|^2 \,\rangle}.
\nonumber
\eeq
for any real $\lambda$. The maximum on the RHS can be minimised over $\lambda$ to give the best energy stability result. 
Clearly, $\rho=0$ is a feature of  the optimiser and 
we can rescale $u$ by defining $\newU:=-\lambda u$ to get
an expression for the energy stability Reynolds number $Re_E$ as
\begin{align}
\dfrac{1}{Re_E} \,:=\, \min_{\lambda} &\dfrac{\lambda^2-2 \Omega(\lambda^2-1)}{\lambda}
\,\nonumber\\
&\times \text{max}_{\newU,v,w}\, 
 \dfrac{\langle \newU v \rangle}
     {\langle  |\bnab \newU|^2+| \bnab v|^2 +|\bnab w|^2 \rangle}
\nonumber
\end{align}
where the implication is that {\em all} streamwise disturbances decay for $Re < Re_E$ regardless of their amplitude.
The latter maximisation corresponds to $1/Re_E$ for an unstratified, non-rotating layer where 
$Re_E=\half \sqrt{1708}\approx 20.7$ \cite{JO70} under non-slip conditions. The minimisation problem has the minimum $2\sqrt{2 \Omega(1-2 \Omega)}$ for $0 \leq \Omega \leq \half$ and $0$ otherwise for real $\lambda$. As a result, we have
\beq
Re_E= \left\{ \begin{array}{ll}
\dfrac{\sqrt{1708}}{4 \sqrt{2 \Omega(1-2\Omega)}} & 0 < \Omega <  \half\\
\infty                                           & \Omega \leq0 \,\,\,{\rm or} \,\,\,\Omega \geq \half
\end{array}
\right.
\nonumber
\eeq 
So, on and beyond the Rayleigh line $\Omega=\half$, the (generalised) energy of all streamwise-independent disturbances, {\it regardless of their amplitude}, monotonically decays in time {\em for any} $Re$. To guarantee that $E_\lambda \rightarrow 0$ (and hence the ultimate vanishing of all disturbance fields), we need a Poincar\'{e} inequality $E_\lambda < \alpha D_\lambda$ for some $\alpha=\alpha(L)$ so that (\ref{E_eqn}) becomes $\mathrm{d}E_\lambda/\mathrm{d}t < -\beta^2 E_\lambda$ for some constant $\beta$. Gr\"{o}nwall's inequality then gives the  required result.
 A Poincare inequality exists for non-slip conditions on the velocity field and either Dirichlet or Neumann conditions on the density field (in the latter case only if no mean flow is allowed 
in the direction of gravity). (Note that once $Re=177.2$ for any $\Omega$, 2D {\it spanwise-invariant} disturbances are not assured to decay \citep{HU72} so that there is no general global stability result for the basic state beyond the Rayleigh line.)


\bibliography{biblio_CI_SRI}

\end{document}